KIU
Journal of Science, Engineering and Technology

# Statistical LOS/NLOS Classification for UWB Channels

Mohammed Dahiru Buhari[1,2], Tri Bagus Susilo[3], and Irfan Khan[4], and Bashir Olaniyi Sadiq[1,5]

*Department of Electrical, Telecommunication and Computer Engineering, Kampala International University, Uganda[1]*
*Electrical and Electronic Engineering Department, Abubakar Tafawa Balewa University, Bauchi, Nigeria[2]*
*Electrical Engineering Department, American University of the Middle East (AUM)[3,4]*
*Department of Computer Engineering, Ahmadu Bello University, Zaria[5].*
buharim@kiu.ac.ug[1], dbmohammed@atbu.edu.ng[2], 3bagussusilo@gmail.com[3], bosadiq@kiu.ac.ug[5]

*Corresponding Author: buharim@kiu.ac.ug[1]*

**Abstract**
*Ultrawideband (UWB) technology has attracted a lot of attention for indoor and outdoor positioning systems due to its high accuracy and robustness in non-line-of-sight (NLOS) environments. However, UWB signals are affected by multipath propagation which causes errors in localization. To overcome this problem, researchers have proposed various techniques for NLOS identification and mitigation. One of the approaches is statistical LOS/NLOS classification, which uses statistical parameters of the received signal to distinguish between LOS and NLOS channels. In this paper, we formulated several techniques which can be used for effectively classifying Line of Sight (LOS) channel from a Non-Line of Sight (NLOS) channel. Various parameters obtained from Channel Impulse Response (CIR) like Skewness, Kurtosis, Root Mean Squared Delay Spread (RDS), Mean Excess Delay (MED), Energy, Energy Ratio and Mean of Covariance Matrix are used for channel classification. In addition to this, the Joint Probability Density Functions (PDFs) of various parameters are used to improve the accuracy of UWB LOS/NLOS channel classification. Two different criteria-Likelihood Ratio and Hypothesis Test are used for the identification of channel.*

**Nomenclature and units**

| | |
|---|---|
| $k$ | Kurtosis |
| $\gamma$ | Skewness |
| $\tau_{MED}$ | Mean Excess Delay |
| $\tau_{RDS}$ | Root Mean Squared Delay Spread |
| $\mu$ | Mean |

## 1.0     Introduction

Ultra-wideband (UWB) propagation presents promising technology for future communication networks due to higher temporal resolution, low power and robust operation in harsh environments. However, obstacles like walls, buildings, vehicles, mountains and trees pose a significant challenge for location estimation as they can result in a positive bias in distance calculation (Marano, Gifford, Wymeersch, & Win, 2010). The next generation devices will not only require the knowledge of channel state but also the knowledge of the exact environment identification (Mucchi, Re, & Landi, 2011). These precise range and location estimates can be made in UWB networks. UWB signals consist of extremely short duration pulses (high temporal resolution) therefore the time of arrival of the receiving signal can be estimated accurately provided that the arrival path can be correctly identified.

In order for a communication system to function effectively under challenging conditions, it must be able to overcome significant obstacles that affect precise ranging and localization. Therefore, distinguishing between (LOS) and Non-Line of Sight (NLOS) signals is crucial, particularly in high resolution positioning systems like UWB systems. To accomplish this, signal classification methods, such as parametric (e.g., Skewness, Kurtosis, and Energy) or non-parametric techniques (e.g., Least square method) can be used (Barral, Escudero, & García-Naya, 2019). To gain a better understanding of the current state of research, it is necessary to examine past literature. In other words, knowledge of the past is essential to comprehend the present and predict the future.

Many researchers have tried to separate the channels that have direct Line of Sight (LOS) from those that do not (NLOS). Venkatesh and his team (Abou-Shehada et al., 2021) suggested a method to detect NLOS channels in UWB systems based on the statistics of the received signal. They used Time of Arrival (TOA), Root Mean Squared Delay Spread (RDS), and Received Signal Strength (RSS), and achieved a clear distinction between LOS and NLOS channels.

Another advancement in this field (Iqbal, Al-Dharrab, Muqaibel, Mesbah, & Stüber, 2020) involved the introduction of Kurtosis as a means of distinguishing between LOS and NLOS in indoor UWB environments. However, this parameter was found to be less effective in outdoor settings. Stefano and colleagues (F. Wang, Tang, & Chen, 2023) developed an algorithm utilizing machine learning techniques to determine whether a signal was transmitted in an LOS or NLOS environment. This method helped to lower the errors in measuring distances in NLOS situations. They also tried another experiment using the Energy, Rise Time, and Maximum Amplitude of the signal they received, which did not need to build specific statistical models. These techniques seem promising for identifying NLOS channels by measuring these parameters.

The authors in (Maranò, Gifford, Wymeersch, & Win, 2010) suggested a new and original method to find and reduce NLOS effects in UWB localization systems, using metric parameters such as Kurtosis, Mean Excess Delay (MED), and RDS. They managed to get a 90% correct identification rate for LOS/NLOS in most of the channel models. In (Kolakowski, 2020) utilized the Kurtosis of the received signal as a parameter for distinguishing between LOS and NLOS channels. This approach not only effectively differentiated between the two, but also enabled the ordering of signal quality in two separate LOS or NLOS rooms, even when the SNR was the same.

Another method was to use joint Time of Arrival (TOA) estimation and NLOS identification based on UWB energy detection (Huang et al., 2023), with the Energy-Based TOA Estimation (EBE) algorithm. The NLOS detection process was done by finding the ratio of the First Path Power (FPP) to the Total Signal Power (TSP) of the Received Signal Folded Version (RSFV) of the EBE algorithm. This method was straightforward and did not require probability distribution multiplication in the identification process. Joon-Young Lee and colleagues (Tabaa, Diou, Saadane, & Dandache, 2014) proposed a hypothesis to determine the presence of LOS blockage in UWB propagation, taking a different approach. After finding out the NLOS channel state, Maximum Likelihood Estimation (MLE) was used for positioning based on TOA. Muqaibel and his team (Iqbal et al., 2020) did a practical test of NLOS/LOS parametric classifications in UWB channels by using parameters like Kurtosis, Peak to Lead Delay (PLD), Mean Excess Delay (MED), and Root Mean Squared Delay Spread (RDS) (Park, Nam, Choi, Ko, & Ko, 2020). Their method successfully found LOS/NLOS with both simulation and experimental data. In (Landolsi, Muqaibel, & Almutairi, 2016), the author examines non-parametric learning algorithms for detecting LOS/NLOS conditions, including Support Vector Machine (SVM), Least Square, and Least Square Support Vector Machine (LS-SVM). The study concludes that non-parametric techniques outperform parametric approaches in terms of efficiency. Similarly, in (Li & Wu, 2012), a non-parametric approach called the chi-square test was utilized for NLOS identification, and two different techniques were proposed for distance estimation: a combination of non-parametric Probability Density Function (PDF) and Kullback Leibler distance, and a maximum likelihood ratio test. In this paper, we use Skewness, Kurtosis, RDS, MED and Energy for channel clarification. We also introduce new parameters such as Energy Ratio, Mean of Covariance Matrix and Joint probability distribution function.

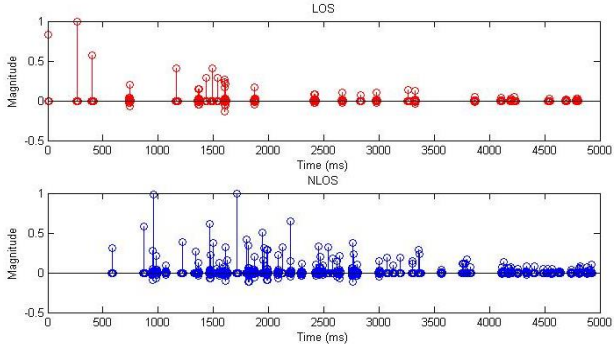

**Figure 1** CIR (a) LOS and (b) NLOS.

The remaining part of the paper is organized as follows: Section III details the methodology employed in this research, while Section IV presents the simulation results. Finally, in Section V, we summarize our conclusions and recommendations for future work.

## 2.0 Materials and Methods

In this section, the materials and methods for this research are presented. Starting from the channel, it can be completely described by its impulse response. In the UWB communication system information is sent via very narrow pulses over the channel. At the receiver end, these pulses can be approximated by the channel impulse response. As can be seen from Fig. 1, the Channel Impulse Response (CIR) is distinct for LOS and NLOS scenarios. This difference in CIR for LOS/NLOS can be exploited for channel state identification. Various parameters calculated from CIR like Skewness, Kurtosis etc. can be utilized to statistically distinguish between LOS/NLOS propagation channel.

### 2.1 Skewness

Skewness is a statistical measure that describes the degree of asymmetry present in a probability density function. Specifically, it indicates the degree to which the probability density of a given random variable is inclined towards one side of the mean over the other. This measure can be mathematically defined as (Truex, Liu, Gursoy, Wei, & Yu, 2019; Y. Wang, Wu, & Cheng, 2018):

$$\gamma = E\left[\frac{(h(t)-\mu)^3}{\sigma}\right] = \frac{\mu_3}{\sigma^3} \quad (1)$$

Where $\gamma$ is the skewness, $\mu_3$ is the third moment about the mean $\mu$, and $\sigma$ is the standard deviation.

Skewness is a parameter that measures the degree of asymmetry of a probability density function. It indicates the extent to which the distribution of a given random variable leans towards one side of the mean. A negative skewness value suggests that the tail on the left side of the PDF is longer than the right side, whereas a positive value implies that the right side is greater than the left side. This property can be used to differentiate between LOS and NLOS channels. When the distribution is symmetric, the skewness value is zero. In the case of LOS channels, most of the peaks are on the left side of the mean, while NLOS signals are distributed more to the right of the mean due to multi-path components.

### 2.2 Kurtosis

The measure of kurtosis for a signal is the ratio of the fourth central moment to the square of its variance, and can be expressed mathematically as follows (Y. Wang et al., 2018):

$$k = \frac{1}{\sigma^4 T}\int_T \left(|x(t)-\mu_{|x|}|\right)^4 dt \quad (2)$$

where

$$\mu_{|x|} = \frac{1}{T}\int_T |x(t)|dt$$

$$\sigma_{|x|}^2 = \frac{1}{T}\int_T \left(|x(t)|-\mu_{|x|}\right)^2 dt$$

Kurtosis is a measure of the peakedness of a distribution. The LOS data normally have a high value of kurtosis compared to NLOS because NLOS signals have high variance due to large number of multiple path components.

### 2.3 MED and RDS

Both MED and RDS show the temporal characteristics of the CIR. MED is given by (Qing, Wei, & Wanchun, 2018):

$$\tau_{MED} = \frac{\int_{-\infty}^{\infty} t|x(t)|^2 dt}{\int_{-\infty}^{\infty} |x(t)|^2 dt} \quad (3)$$

RDS is given by:

$$\tau_{RDS} = \sqrt{\frac{\int_{-\infty}^{\infty}(t-\tau_M)^2|x(t)|^2}{\int_{-\infty}^{\infty}|x(t)|^2 dt}} \quad (4)$$

The two parameters mentioned provide an indication of how the energy of the received signal is dispersed. In a LOS environment, the energy is more focused on the initial path, while in an NLOS environment, it is more evenly spread over multiple multipath components (Tian, Wei, Wang, & Zhang, 2019).

### 2.4 Mean of Covariance Matrix

The degree to which LOS and NLOS scenarios vary together will be used to identify the incoming data. Covariance matrix is given as (Zeng, Chang, Zhang, Hu, & Li, 2018):

$$\Sigma_{ij} = E[(h_i - \mu_i)(h_j - \mu_j)] \quad (5)$$

where
$\Sigma_{ij} = cov(h_i, h_j)$
$E[*] = Expected\ value$
$\mu = Mean$

Indices i and j represent test and reference data (LOS or NLOS scenarios). By taking the mean of every column of the covariance matrices (LOS and NLOS reference data), we can classify the test data to be LOS or NLOS.

### 2.5 Energy and Energy Ratio

Total energy of the signal can be used for identification of LOS/NLOS scenario. NLOS signal suffers considerable energy losses due to collision and reflections during propagation. The energy of NLOS signal is much smaller as compared to LOS. This difference in energy can be exploited for LOS/NLOS channel identification (Tekbıyık, Tokgöz, Ekti, Yarkan, & Kurt, 2020).

Energy ratio is defined as the ratio between the first non-zero samples of the receiving signal to the energy of the signal. These parameters can be calculated using:

$$\varepsilon_\tau = \int_{-\infty}^{+\infty} |r(t)|^2 dt \quad (6)$$

$$\varepsilon_r = \frac{\varepsilon_1}{\varepsilon_\tau} \quad (7)$$

where $\varepsilon_\tau$ is the total energy of the received signal, $\varepsilon_1$ is the energy of the first nonzero sample, and $\varepsilon_r$ is the energy ratio.

Since LOS signals have most of their energy on the first arrival, this ratio is high for LOS signals as compared to NLOS signals.

### 2.6 Joint Density Functions

Joint density between various parameters like RDS, MED, Kurtosis etc. can be used to increase the accuracy of LOS/NLOS identification. For joint density function, PDF of all the parameters is assumed to be independent of each other. To numerically evaluate the accuracy of LOS/NLOS classification we develop two different tests i.e., Likelihood Ratio Test and Hypothesis Test. These tests are discussed as follow:

### 2.6.1 Likelihood Ratio Test

For a specific parameter the likelihood ratio test is defines as

$$\frac{P_{LOS}(k)}{P_{NLOS}(k)} \underset{H_1}{\overset{H_0}{\gtrless}} 1 \quad (8)$$

To classify a given channel as either LOS or NLOS, the probability ratio of each is computed. If the ratio is greater than 1, the channel is labeled as LOS; otherwise, it is labeled as NLOS. In the case of joint probability density function, the Likelihood Ratio test is utilized, as follows (Ding et al., 2016):

$$\frac{P_{LOS}(k_1, k_2, \ldots k_n)}{P_{NLOS}(k_1, k_2, \ldots k_n)} \underset{H_1}{\overset{H_0}{\gtrless}} 1 \quad (9)$$

The simplification of the Ratio Test for joint distribution can be achieved by assuming that the parameters are statistically independent. The Ratio Test for the second-order joint distribution can be expressed as follows:

$$\frac{P_{LOS}(k_1, k_2)}{P_{NLOS}(k_1, k_2)} = \frac{P_{LOS}(k_1)}{P_{NLOS}(k_1)} \times \frac{P_{LOS}(k_2)}{P_{NLOS}(k_2)} \underset{H_1}{\overset{H_0}{\gtrless}} 1 \quad (10)$$

### 2.6.2 Hypothesis Test

In this test, the received signal parameters like kurtosis, PLD etc. are compared with a pre-defined threshold, based on which a decision is made about the channel state. Selecting a specific value of threshold, channel state can be identified as either LOS or NLOS (D. Wang & Tellambura, 2020).

If the value of the parameter is greater than the threshold T, the path is classified as LOS, while if it is less than T, it is classified as NLOS, or vice versa.

$$Test: \begin{cases} LOS, & if\ k \leq Threshold \\ NLOS, & if\ k > Threshold \end{cases} \quad (11)$$

The Ratio Test and Hypothesis Test give the same result if there is only a single point of intersection between the PDF of LOS and NLOS channel realizations. In the case of multiple intersections, the Ratio Test and Hypothesis Test produce different results.

## 3.0 Results and Discussions

In this section, we present our analysis for the parametric approach in identifying the UWB channel in LOS/NLOS channel. The PDFs for various parameters like Skewness, Kurtosis, RDS, MED and Mean of Covariance Matrix were used for channel classifications as well as their joint PDFs. Also, the ratio test and the hypothesis test were used to get numerical results for the accuracy of each channel state.

Each of these parameters are used to evaluate each LOS/NLOS channel realization. Data used in this paper consists of 1000 channel realization of LOS and NLOS each from 0 to 10 seconds with time resolution of 0.02 seconds.

In Fig. 2, PDF for Skewness is shown for both LOS and NLOS scenario. The accuracy of identifying a channel as LOS or NLOS is 70.6% and 68.4% respectively using the Ratio Test. For Hypothesis Test, the threshold is taken at 18.1506, and the accuracy for LOS/NLOS is 72.9% and 68.4% respectively.

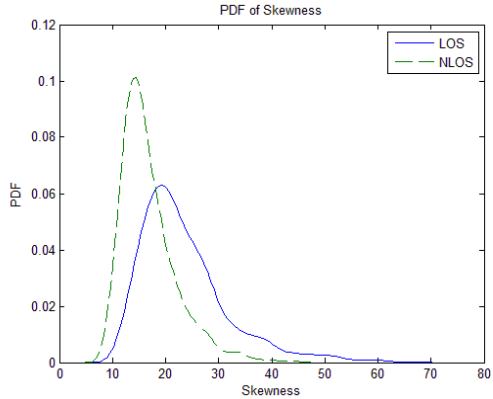

**Figure 2** PDF of Skewness.

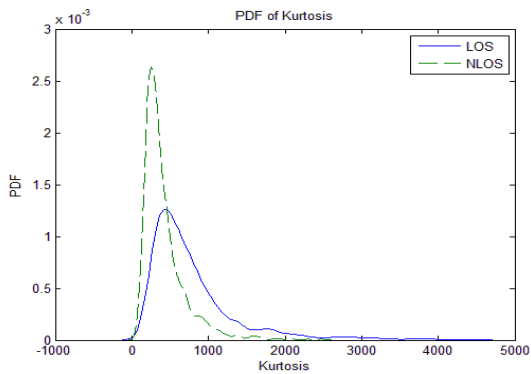

**Figure 3** PDF of Kurtosis.

In Fig. 3, PDF for Kurtosis is shown for LOS/NLOS scenario. By using this parameter, the accuracy of identifying a channel as LOS/NLOS is 66.6% and 70.9% respectively using the Ratio Test. For Hypothesis Test, the threshold is taken at .0021, and the accuracy for LOS/NLOS is 73.1% and 66.3% respectively.

PDF of RDS and MED as shown in Fig. 4 and Fig. 5 represents the temporal characteristics of the CIR. Maximum accuracy for RDS is 50.6% for LOS and 82.4% for NLOS channel using Ratio test. For the Hypothesis test, we chose our threshold at 14.205. The results for Hypothesis test for RDS are 45.4% and 85.3% for LOS and NLOS respectively.

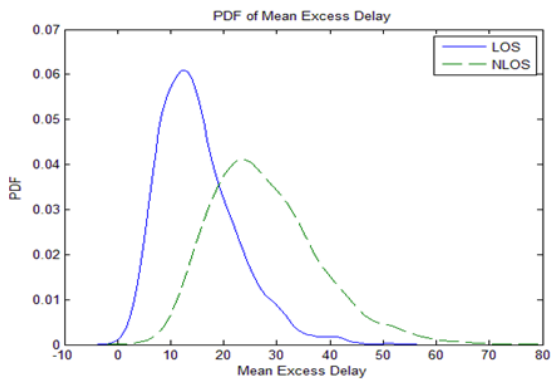

**Figure 4** PDF of MED.

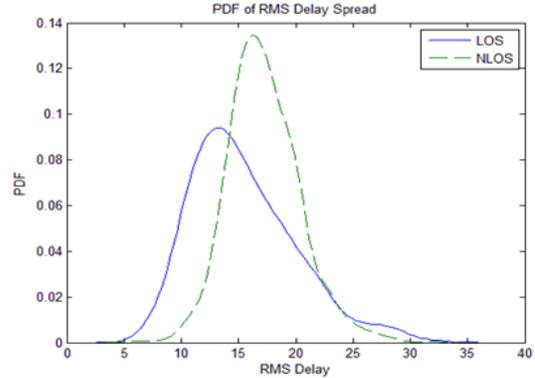

**Figure 5** PDF of RDS.

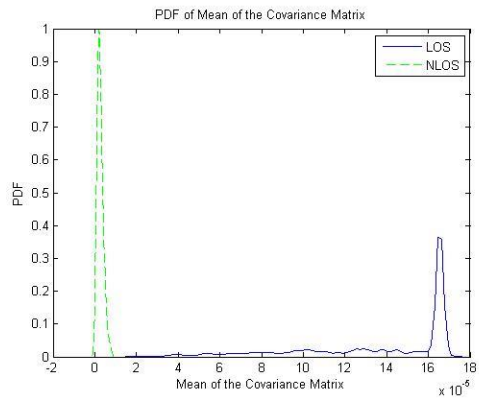

**Figure 6** PDF of Mean of Covariance.

The PDF of MED provides the best result for classification of UWB channel for our current realization. The Ratio test results are 75% for LOS and 79% for NLOS. For Hypothesis test, we choose our threshold at 19.1104 and the accuracy for LOS/NLOS identification is 75% and 80.2% respectively.

In Fig. 6, PDF of Mean of Covariance Matrix is shown. It informs that there is no intersection between LOS and NLOS density functions. In other words, LOS and NLOS channels can be separated completely.

PDF for Energy and Energy Ratio are shown in Fig. 7. Accuracy of LOS/NLOS classification for Energy is 67.8% and 66.3% using the Ratio test and 67.8% for both LOS/NLOS using the Hypothesis Test with threshold is taken at 5.0031.

PDF of Energy Ratio is taken between the first arrival and the total signal energy. Since in NLOS the first arrival is usually very small as compared to LOS in which the first arrival is usually the peak value, this feature can be exploited to identify LOS/NLOS channel state. It can be seen from Figure 7 that both the PDF's are separated at the threshold of .0273. Using the Energy Ratio, the accuracy for LOS/NLOS is 94.6% and 99.7% respectively.

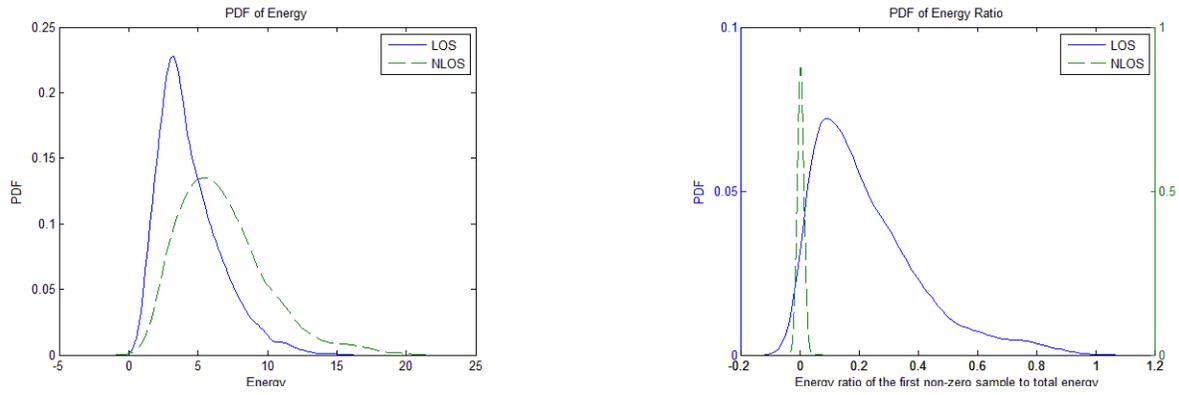

**Figure 7** PDF of Energy and Energy Ratio.

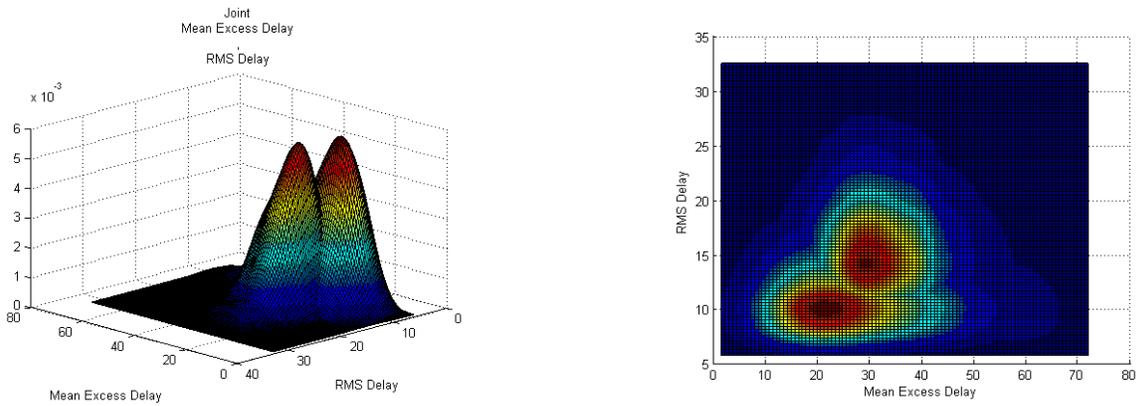

**Figure 8** Joint Density of MED and RDS.

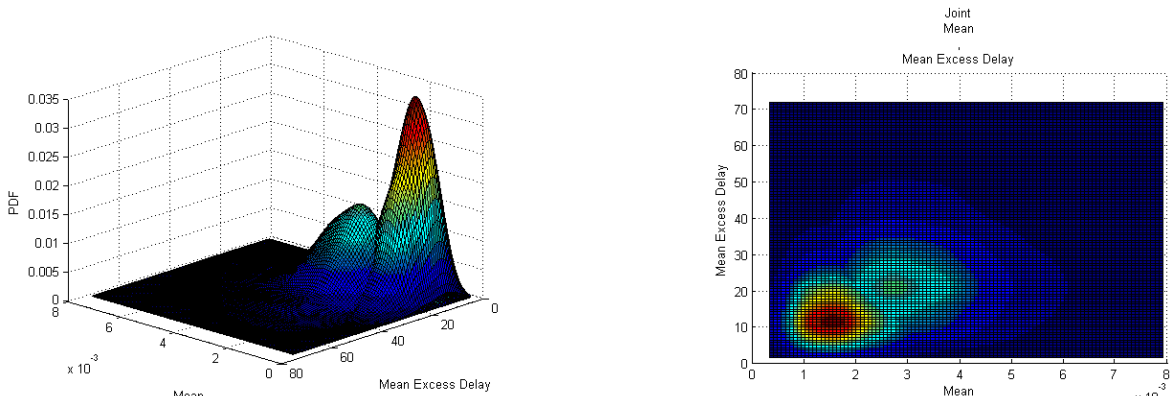

**Figure 9** Joint Density of Mean and MED.

Fig. 8 shows the joint density function between MED and RDS. The ratio test results are 58.07% for LOS and 89.84% for NLOS. Fig. 9 shows the joint probability density function between Mean and MED. The left part of Fig. 9 illustrates that the density function of LOS is highly concentrated on the right corner, represented by a tall peak. On the other hand, the NLOS data is represented by a wider peak, which is smaller in size. Table I summarizes all the results discussed in this section.

**Table 1** LOS/NLOS Classification Percentages.

| Parameter | Ratio Test | | Hypothesis test | | |
|---|---|---|---|---|---|
| | *LOS* | *NLOS* | *LOS* | *NLOS* | *Threshold* |
| Skewness | 70.6 | 68.4 | 72.9 | 68.4 | 18.1506 |
| Kurtosis | 66.6 | 70.9 | 73.1 | 66.3 | .0021 |
| Energy | 67.8 | 66.3 | 67.8 | 67.8 | 5.0031 |
| Energy Ratio | - | - | 94.6 | 99.7 | .0273 |
| RMS delay | 50.6 | 82.4 | 45.4 | 85.3 | 14.205 |
| MED | 75 | 79 | 75 | 80.2 | 19.11 |
| Mean of Covariance Matrix | - | - | 100 | 100 | 1.2237e-05 |
| $\tau_{rms}$ & $\tau_{med}$ | 58.07 | 89.84 | - | - | - |
| Mean & $\tau_{med}$ | 75.05 | 70.71 | - | - | - |

## 5.0 Conclusions

In this paper we investigated several parametric techniques which can be used for statistically classifying LOS path from an NLOS path. Various parameters obtained from channel impulse response like Kurtosis, Skewness, RDS, MED, Energy, Energy Ratio, Mean of Covariance Matrix and the joint PDF of various parameters are used for channel identification. Mean of Covariance Matrix and Energy Ratio give the best result among all the parameters in classification between LOS/NLOS.

## Acknowledgements
The authors would like to express their gratitude to the King Fahd University of Petroleum and Minerals, KSA and the Faculty of Engineering and Applied Science, KIU for the support provided while carrying out this research.

## Declaration of conflict of interest

The authors have collectively contributed to the conceptualization, design, and execution of this journal. They have worked on drafting and critically revising the article to include significant intellectual content. This manuscript has not been previously submitted or reviewed by any other journal or publishing platform. Additionally, the authors do not have any affiliation with any organization that has a direct or indirect financial stake in the subject matter discussed in this manuscript.